\documentstyle[aps]{revtex} 

\input epsf

\begin{document}
\twocolumn[
\title{Approach to Perturbative Results in the $N-\Delta$ Transition}

\author{ Carl E. Carlson }
\address{ Physics Department, College of William and Mary, Williamsburg, VA
23187-8795 }

\author{ Nimai C. Mukhopadhyay }
\address{ Department of Physics, Applied Physics, and Astronomy,  Rensselaer
Polytechnic Institute,  Troy,  NY 12180-3590 }
\date{April 21, 1998}
\maketitle
\parshape=1 0.75in 5.5in \indent {\small {We show that constraints from
perturbative QCD calculations play a role in the nucleon to Delta(1232)
electromagnetic transition even at moderate momentum transfer scales.  The
pQCD constraints, tied to real photoproduction data and unseparated resonance
response functions, lead to explicit forms for the helicity amplitudes
wherein the E2/M1 ratio remains small at moderately large momentum transfer.}
 
\vglue -20pt}

\widetext \vglue -5.9cm  \hfill RPI-98-N126; WM-98-105; hep-ph/9804356
\vglue 5.4cm \narrowtext
\pacs{13.88+e, 14.20.Gk, 13.60.Hb, 12.40.N, 12.38.Bx}

]




The nucleon-Delta(1232) electromagnetic transition at very low momentum transfer
is mainly a magnetic dipole (M1 or $M_{1+}$) transition, with a small electric
quadrupole (E2 or $E_{1+}$) component in addition~\cite{dm97}.  The so called
electromagnetic ratio (EMR), which is the ratio $E_{1+}/M_{1+}$ at the Delta
peak, is a few percent in magnitude and negative in this region.  In contrast,
at very high momentum transfer, perturbative QCD (pQCD) should be valid and
demands that this ratio approach unity~\cite{c86}.  How should the M1 and E2
amplitudes interpolate between these two extremes?  In particular, can the pQCD
results be in any way relevant at momentum transfers experimentally reachable
now or in the near future?  That is the concern of this paper.

One clear issue is that the existing experiments, with momentum transfers up to
3.2 GeV$^2$~\cite{haiden}, do not show any hint that E2 is becoming positive and
large, when one examines what are considered the most sophisticated
analyses~\cite{analyses}.   The two analyses of ref.~\cite{analyses} 
give $E_{1+}/M_{1+} =0.06 \pm 0.02 \pm 0.03$ and $0.0\pm 0.14$, respectively, at
3.2 GeV$^2$. Indeed, there is a school of thought holding that the pQCD limit is
not experimentally reachable in exclusive reactions~\cite{ils89}.  

We shall here examine the point of view that pQCD results can be relevant to
moderate momentum transfer squared ($q^2$) exclusive reactions, and that
considering the approach to the pQCD limit can be useful in understanding lower
$q^2$ data being studied at present.  We shall interpolate between the very low
and very high
$q^2$ domains using analytic functions motivated by simplicity, and using what
is known about the $q^2=0$ point and the asymptotic pQCD limit as anchor
points.  The latter includes the known scaling laws~\cite{count,c86} and
normalized leading twist calculations~\cite{cp88,fozz89,bone}, where
possible.  Future data will throw further light on the validity of our
interpolations.

Our choice to study the Delta transition amplitudes is motivated by their
special characteristic.  Delta electromagnetic production falls relative to
continuum with increasing $q^2$, in contrast to other resonances where the
resonance signal relative to background is roughly constant.  This is mirrored
in the normalized pQCD calculation of the leading twist helicity amplitudes,
which show that asymptotic N-Delta transition amplitude is anomalously small.




Before showing our interpolations, let us review the notations.  The transverse
electromagnetic helicity amplitudes $A_{1/2}$ and $A_{3/2}$ for the N-Delta
transition are related to the multipole amplitudes by~\cite{dmw91}
 
\begin{eqnarray}  
\label{convert} M1 = -{1\over 2} A_{1/2} - {\sqrt{3}\over 2}
A_{3/2},  \nonumber \\ E2 = -{1\over 2} A_{1/2} + {1\over 2\sqrt{3}} A_{3/2}.
\end{eqnarray}
At very high $Q^2 = - q^2$, pQCD predicts the scaling behavior of the helicity
amplitudes to be~\cite{c86,cp88}

\begin{eqnarray} 
A_{1/2} \propto {1\over Q^3} \ ,          \nonumber \\ 
A_{3/2} \propto {1\over Q^5} \ ,
\end{eqnarray}
modulo $\log(Q^2)$ factors.  Hence, the asymptotic 
($Q^2 \rightarrow \infty$ )prediction that $E2/M1 \rightarrow 1$ follows.   

We have already mentioned that the asymptotic N-Delta transition amplitude is
small.  In order to realize how small the asymptotic N-Delta transition
amplitude actually is~\cite{stoler,cm93}, one should quote the electromagnetic
transition amplitudes for the elastic and quasielastic cases in comparable
fashions.  One can translate the leading amplitude in all cases into the
helicity amplitude $G_+$,

\begin{equation} 
G_+ = {1\over 2m_N} \langle R, \lambda^\prime = {1\over 2} |
   \epsilon_\mu^{(+)} \cdot j^\mu(0) | N, \lambda = {1\over 2} \rangle .
\end{equation}
where $\epsilon_\mu^{(+)}$ is the photon polarization vector for photon
helicity +1, $j^\mu$ is the electromagnetic current, $\lambda$ refers to
helicity, and the overall mass factor is included to make $G_+$ dimensionless. 
Asymptotically, $G_+$ scales the same way as $A_{1/2}$.

For the elastic case we relate $G_+$ to $G_M$ by

\begin{equation} 
Q^3 G_+(p \rightarrow p) = {1\over m_N \sqrt{2}} Q^4 G_M 
           \approx 0.75 {\rm \ GeV}^3  ,
\end{equation}
where the last part is valid at large $q^2$ and the numerical value comes both
from data and from calculations using any of the standard nucleon distribution
amplitudes mentioned below.  For nucleon-resonance transitions, the result is
most commonly quoted in  terms of the helicity amplitude
$A_{1/2}$~\cite{cko69,warns90} which has some factors of charge and momentum
multiplied in.  In terms of $G_+$,

\begin{equation} 
Q^3 G_+(N \rightarrow R)
    = {1\over e} \sqrt{m_R^2 - m_N^2 \over m_N} Q^3 A_{1/2}.
\end{equation}
where $e$ is the elementary charge.

For the Delta(1232), the calculations show a small asymptotic
$A_{1/2}$ and in terms of $G_+$

\begin{equation}  \label{asymp} 
Q^3 G_+(N \rightarrow \Delta) = \left\{
\begin{array}{cl} 0.05 {\rm GeV}^{5/2}  &  \quad {\rm CZ} \\ 0.08 {\rm
GeV}^{5/2}  &  \quad {\rm KS \,,}
\end{array}
\right.  
\end{equation}
where the calculations used the Delta distribution amplitude calculated in QCD
sum rule calculations in Ref.~\cite{cp88} and the nucleon distribution
amplitude calculated by either Chernyak and Zhitnitsky (CZ,~\cite{cz}) or King
and Sachrajda (KS,~\cite{ks})~\cite{gs}.   The uncertainties in the QCD sum
rule determination of the Delta distribution amplitude are sufficient that the
correct answer could be two or three times larger or smaller than the above
quoted results.  Nonetheless, the leading N-Delta transition amplitude appears
to be truly small.  This is underscored by comparing to the N-N*(1535)
transition, for which the normalized pQCD calculation is also
possible~\cite{cp88} and leads to

\begin{equation} 
Q^3 G_+(p \rightarrow N^*(1535)) = \left\{
\begin{array}{cl} 
0.46 {\rm GeV}^{5/2}  &  \quad {\rm CZ} \\ 
0.58 {\rm GeV}^{5/2}  &  \quad {\rm KS \,.}
\end{array}
\right.
\end{equation}

 This brings us to our main question: what functional form shall we choose to
interpolate between the low $q^2$ and asymptotic domains?  We can receive
guidance from the nucleon elastic case.  The helicity amplitude $G_+$for the
nucleon has a kinematic zero at $Q=0$, and after noting that 
$G_+ \propto QG_M$, it is known that the magnetic from factor $G_M$ is decently
fit with a dipole form.  For resonance production the kinematic zero moves to
the ``pseudothreshold'' or ``no-recoil'' point, where in the resonance rest
frame the nucleon is also at rest.  This point is the threshold for Dalitz
decay of the resonance, $R \rightarrow N+ \gamma^* \rightarrow N e^+ e^-$. 
The kinematic zero is proportional to powers of $|\vec q\,^*|$, the  momentum
of the photon in the resonance rest frame, and the number of powers is one for
both $G_\pm$ in
$\Delta$ electroproduction~\cite{bw66}.  Further, both the nucleon and Delta
are in the ground state 56-plet of the approximate SU(6) spin-flavor symmetry,
so we feel it is a good {\it Ansatz} to also using a simple dipole form for the
$A_{1/2}$ amplitude for the $N$-$\Delta$ transition, and a similar form with
one more asymptotic power of $Q^{-2}$ for $A_{3/2}$.  The controlling factor in
$|\vec q\,^*|$ is~\cite{extrazeronote}

\begin{equation} 
Q^* = \sqrt{Q^2 + (m_R - m_N)^2}.
\end{equation}
Hence we shall take our interpolating forms to be

\begin{eqnarray} 
A_{1/2}(Q^2) = {Q^*\over m_\Delta - m_N}
     {A_{1/2}(0) \over ({1+{Q^2/ \Lambda_1^2})^2}}\ ,\nonumber \\ 
A_{3/2}(Q^2) =
{Q^*\over m_\Delta - m_N}
     {A_{3/2}(0) \over ({1+{Q^2/ \Lambda_3^2})^3}}\ ,
\end{eqnarray}
where $\Lambda_1$ and $\Lambda_3$ are parameters.

At low $q^2$ the multipole amplitudes are more natural, and one expects as well
as sees a dominance of the M1 amplitude.  If the dominance is complete, one
expects from Eq.~(\ref{convert})

\begin{equation} 
A_{3/2}(0) = \sqrt{3} A_{1/2}(0).
\end{equation}
Since E2 is not quite zero, we shall use real photon helicity amplitudes given
by the data~\cite{dm97}.   

At the high $q^2$ end, we have

\begin{equation}
\lim_{Q^2 \rightarrow \infty} Q^3 A_{1/2} = A_{1/2}(0) 
           {\Lambda_1^4 \over m_\Delta - m_N}.
\end{equation}
The parameter $\Lambda_1$ can now be constrained by the calculated asymptotic
values of the left hand side, Eq.~(\ref{asymp}).  The data may indicate a
somewhat different value.  In any case, since $Q^3 A_{1/2}$ is asymptotically
small, we have reason to expect $\Lambda_1$ to be small compared to the typical
scale exemplified by the mass parameter of the nucleon form factor.  This means
that the helicity amplitude $A_{1/2}(N\rightarrow \Delta)$ will show an {\em
anomalously rapid} falloff.

\begin{figure}

\hskip 0. in \epsfxsize 3.5 in 
\epsfbox{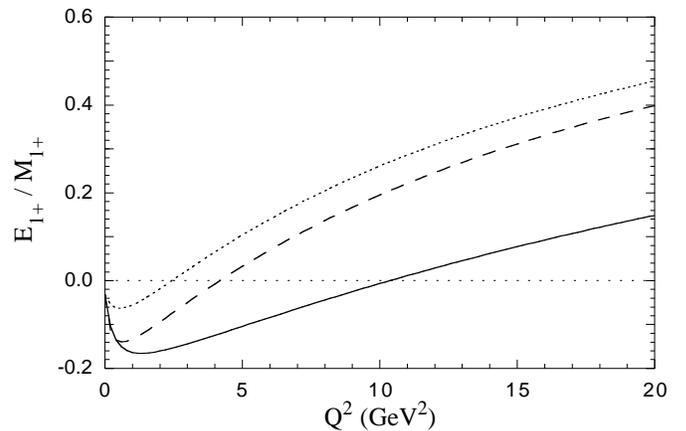}

\hskip 0.1 in

\caption{The E2/M1 ratio for the N-$\Delta$ transition as a function of $Q^2$. 
The dashed curve is the benchmark described in the text, the solid curve is our
preferred parameterization taking into account a number of constraints, and the
dotted curve is another parameterization with a larger asymptotic value of
$A_{1/2}$.  The amplitudes in the last two parameterizations fit the
unseparated data well.}

\label{emr}
\end{figure}

Regarding $\Lambda_3$, we have no special guidance from pQCD but also no reason
to think that $A_{3/2}(N\rightarrow \Delta)$ should be anomalous at high $q^2$. 
So we could choose for example the same value found for the dipole approximation
to the nucleon form factor, suggesting $\Lambda_3^2 = 0.71 {\rm \ GeV}^2$ as a
benchmark.  Alternatively, we could fit $\Lambda_3$ to the intermediate $q^2$
data.  The latter is what we shall do.

Fig.~\ref{emr} illustrates the expected E2/M1 ratio under two specific
assumptions of hadron dynamics. The dotted curve is our benchmark with
$Q^3A_{1/2}$ in the asymptotically large $Q^2$ limit given by the KS amplitudes,
with the parameter $\Lambda_3$ given by the dipole scale of the electromagnetic
form factor of the nucleon. We shall return to the solid curve shortly.  We note
that the falloff of the helicity amplitudes as functions of
$Q^2$ is rapid. This is specially true for $A_{1/2}$, explaining why the EMR
stays negative to several GeV$^2$.

It is here that we can benefit for the existing unpolarized data on the quantity
$G_T$ which is proportional to sum of the squares of the helicity amplitudes,
compiled by Stoler~\cite{stoler},

\begin{equation}
\left| G_T(Q^2) \right|^2 = {2m_N^2 \over Q^2} 
         \left( |G_+|^2 + |G_-|^2 \right)  ,
\end{equation}
and compared to the dipole form,

\begin{equation}
G_{dipole} = {2.79 \over \left( 1+Q^2/0.71 {\rm\, GeV}^2 \right)^2 } \ \ .
\end{equation}

\begin{figure}

\hskip 0. in \epsfxsize 3.5 in 
\epsfbox{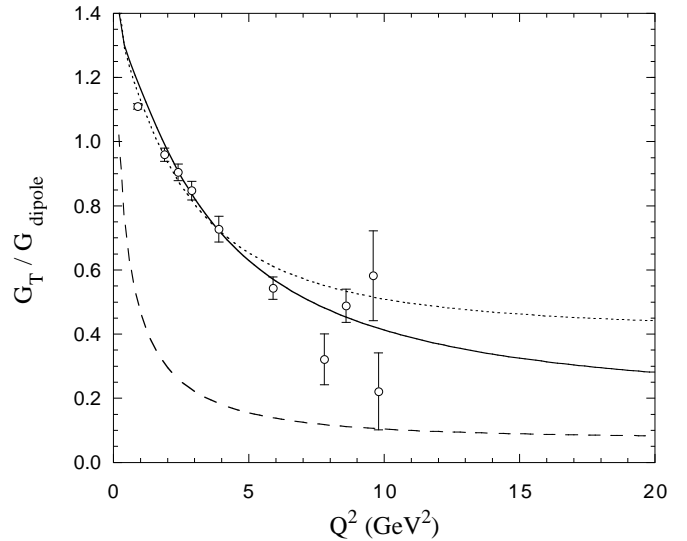}

\hskip 0.1 in

\caption{Our parameterizations compared to the unseparated data for 
$A_{1/2}^2 + A_{1/2}^2$ presented as the quantity
$G_T/G_{dipole}$, defined in the  text.  The curves match Fig. 1 and the data
is from Table 5 of the second of~\protect\cite{stoler}.}

\label{unseparated}
\end{figure}

 Our benchmark described earlier does not quite reproduce this dataset in the
Delta region (Fig.~\ref{unseparated}, dotted curve). A tuning of the 
asymptotic value of $Q^3A_{1/2}$ to 0.08 GeV$^5/2$(or $Q^3G_+$ to 0.22
GeV$^3$), along with the parameter
$\Lambda_3$ adjusted to 1.14 GeV describe this data a lot better, as
illustrated by the solid curve of Fig.~\ref{unseparated}. This forms our more
sound basis for predicting the EMR behavior as a function of $Q^2$. This is
shown by the solid curve in Fig.~\ref{emr}.  Current experiments, under
analysis at CEBAF in the Jefferson Lab, will test  this prediction in the near
future.   

As another possibility, we show in Fig.~\ref{emr} the prediction using a larger
asymptotic value $Q^3A_{1/2} = 0.17 {\rm GeV}^{5/2}$ (which happens to be what
is obtained using the Gari-Stefanis distribution amplitude for the
Delta~\cite{gs,gs86}) and  the $\Lambda_3$ shrunken slightly to 1.10 GeV.  This
change in $A_{1/2}$ has little effect on Fig.~\ref{unseparated} below 5
GeV$^2$.

Considerations of the longitudinal helicity amplitudes are also possible in the
same spirit.  They do require considerations of pseudothreshold
constraints~\cite{js73}, and lie outside the scope of the present paper.

In summary, it is an important question how one should interpolate
between the constituent quark model at low momentum transfers and perturbative
QCD at high momentum transfers.  The nucleon to Delta electromagnetic transition
may be particularly instructive because the asymptotic pQCD prediction for the
EMR is far from what is seen at the highest momentum transfers for which there
are data reported~\cite{haiden}.  One may even entertain the idea that pQCD is
irrelevant at feasibly measurable momentum transfers.  Our considerations lead
to the opposite conclusion.

We take interpolating functions that have the correct behavior near the
no-recoil point, have the correct asymptotic pQCD behavior, are normalized at
the photon point by the constituent quark model or by data, and have an
asymptotic normalization that is guided when possible by normalized pQCD
calculations (tuned by data).  Such functions are simple, plausible, and fit
the data well.  Further, despite the fact that their $Q^2$ dependence and large
$Q^2$ normalization are in accord with pQCD, they lead directly to having the
EMR remain negative to momentum transfers of many, though fewer than 10,
GeV$^2$.

To conclude, our present study in the Delta resonance region has thrown 
light on the role of perturbative physics as constrained in the value of
the quantity $Q^3A_{1/2}$ for asymptotically large $Q^2$ in influencing
the behavior of the E2/M1 ratio for the nucleon to Delta transition. Given
the abnormal suppression of the normally dominant amplitude $A_{1/2}$ in
the case of the Delta excitation, anticipated by the nucleon and Delta
amplitudes inferred from the QCD sum rule approaches, the parameter 
controlling the fall-off of the subleading amplitude $A_{3/2}$ also plays
a crucial role in determining this ratio. New experimental results on this
from electron factories will test this rigorously, and they are eagerly
awaited.

We thank Richard M. Davidson, Valery Frolov, Manfred Gari, Charles Hyde-Wright,
Paul Stoler, and Dirk Walecka for stimulating discussions.  CEC thanks the NSF
for support under grant PHY-9600415; and NCM is grateful to the U. S. Department
of Energy for its support through grant DE-FG02-88ER40448.


\begin{thebibliography}{99}
\bibitem{extrazeronote} The full expression for $|\vec q\,^*|$ is
     $(2m_R)^{-1} \times$
%
     \begin{eqnarray} 
          \left[ \left( Q^2+(m_R-m_N)^2 \right)^2  
                         \left( Q^2+(m_R+m_N)^2 \right)^2 \right]^{1/2},
     \nonumber
     \end{eqnarray}
%
     which might suggest another kinematic zero at another value of $Q^2$. 
     However, the argument that leads to kinematic 
     zeros~\cite{bw66} as powers of 
     $|\vec q\,^*|$ applies only at pseudothreshold, and not at the 
     $\gamma^* \rightarrow R+N$ production threshold.  An easy case to check
     explicitly is an imagined elementary nucleon, where $G_+$ has a factor $Q$
     but no factor $\sqrt{Q^2+4m_N^2}$.
\bibitem{gs} One may recall that a larger N-Delta result can be gotten from the
Gari-Stefanis~\cite{gs86} nucleon distribution amplitude. However, the latter is
now seems to be ruled out because it predicted a much smaller neutron elastic
form factor than has since been seen~\cite{lung93}.
\end{thebibliography}
\end{document}